\documentclass[manuscript]{aastex}
\usepackage{apjfonts}
\usepackage{graphicx}

\newcommand{\Line}[3]{\Ion{#1}{#2}\,$\lambda$\,#3}

\newcommand{\Ion}[2]{#1{\,\scriptsize #2}}

\newcommand{\Mwd}{\mbox{$M_1$}}
\newcommand{\Msec}{\mbox{$M_2$}}

\newcommand{\Porb}{\mbox{$P_{\rm orb}$}}
\newcommand{\Pspin}{\mbox{$P_{\rm spin}$}}

\newcommand{\fullwidth}{\mbox{$\Delta\phi_{\rm 1/2}$}}

\newcommand{\T}{\mbox{$T_\mathrm{0}$}}
\newcommand{\id}{\mbox{$\mathrm{d^{-1}}$}}
\newcommand{\Msun}{\mbox{$\mathrm{M}_{\odot}$}}
\newcommand{\Rsun}{\mbox{$\mathrm{R}_{\odot}$}}

\begin{document}

\title{IPHAS\,J062746.41+014811.3: a deeply eclipsing intermediate polar} 

\author{
A. Aungwerojwit\altaffilmark{1,2},
B.T. G\"ansicke\altaffilmark{3},
P.J. Wheatley\altaffilmark{3},
S.Pyrzas\altaffilmark{3},
B. Staels\altaffilmark{4},
T. Krajci\altaffilmark{5},
and P. Rodr\'iguez-Gil\altaffilmark{6,7,8}
}

\altaffiltext{1}{Department of Physics, Faculty of Science, Naresuan University,
 Phitsanulok, 65000, Thailand}

\altaffiltext{2}{ThEP Centre, CHE, 328 Si Ayutthaya Road, Bangkok, 10400, Thailand}

\altaffiltext{3}{Department of Physics,University of Warwick, Coventry
  CV4 7AL, UK} 
\altaffiltext{4}{CBA Flanders, Alan Guth Observatory, Koningshofbaan
  51, Hofstade, Aalst, Belgium}
\altaffiltext{5}{Astrokolkhoz Observatory, 1351 Cloudcroft, NM 88317, USA}
\altaffiltext{6}{Instituto de Astrof\'isica de de Canarias, V\'ia L\'a actea, s/n, La Laguna, E-38205, Tenerife, Spain}
\altaffiltext{7}{Departamento de Astrof\'isica de, Universidad de La Laguna, Avda. Astrof\'sico Fco. S\'anchez, sn, La Laguna, E-38206, Tenerife, Spain}
\altaffiltext{8}{Isaac Newton Group of Telescopes, Apartado de correos 321, S/C de la Palma, E-38700, Canary Islands, Spain}

\begin{abstract}

We present time-resolved photometry of a cataclysmic variable
discovered in the Isaac Newton Telescope Photometric $\rm{H}\alpha$
Survey of the northern galactic plane, IPHAS\,J062746.41+014811.3 and
classify the system as the fourth deeply eclipsing intermediate polar
known with an orbital period of $\Porb=8.16$\,h, and a spin period of
$\Pspin=2210$\,s. The system shows mild variations of its brightness,
that appear to be accompanied by a change in the amplitude of the spin
modulation at optical wavelengths, and a change in the morphology of
the eclipse profile. The inferred magnetic moment of the white
  dwarf is $\mu_{\rm{wd}}\sim6-7\times10^{33}$\,G\,$\rm{cm^{3}}$, and
  in this case IPHAS\,J0627 will either evolve into a short-period
  EX\,Hya-like intermediate polar with a large $\Pspin/\Porb$ ratio,
  or, perhaps more likely, into a synchronised polar. \textit{Swift}
observations show that the system is an ultraviolet and X-ray source,
with a hard X-ray spectrum that is consistent with those seen in other
intermediate polars. The ultraviolet light curve shows orbital
modulation and an eclipse, while the low signal-to-noise ratio X-ray
light curve does not show a significant modulation on the spin period.
The measured X-ray flux is about an order of magnitude lower than
would be expected from scaling by the optical fluxes of well-known
X-ray selected intermediate polars.
\end{abstract}

\keywords{stars: individual (IPHAS\,J062746.41+014811.3) --
          cataclysmic variables, intermediate polar, eclipsing        
}

\section{Introduction}

Cataclysmic variables (CVs) are semi-detached close binary systems
comprising an accreting white dwarf, and a late-type main-sequence
donor. The strength of the magnetic field of the white dwarf plays an
important role in governing the process of accretion. If the magnetic
field is weak, mass transfer takes place via an accretion disk. In
contrast, if the magnetic field is strong enough ($B\sim10-200$\,MG)
to suppress the formation of the disk, the accretion stream from the
secondary star flows along the magnetic field lines to the poles of
the white dwarf. These systems are known as polars. For moderate
magnetic-field strength systems ($B\sim1-10$\,MG), or intermediate
polars (IPs), the transferred material may form a partial disk in
which the inner part is disrupted into accretion curtains that channel
material to the magnetic poles of the white dwarf. In polars, the
rotational period of the white dwarf ($\Pspin$) is generally
synchronised to the orbital period ($\Porb$), whereas the white dwarfs
in IPs are rotating asynchronously with $\Pspin/\Porb\sim0.01-0.6$
(see \citealt{warner95-1} for a comprehensive review on CVs).

The evolution of magnetic CVs is still subject to
discussion. Observationally, polars and IPs dominate the population of
magnetic CVs below and above the 2--3\,h orbital period gap,
respectively. Both classes overlap in magnetic field strength,
suggesting that IPs with relatively high fields may synchronise once
they have evolved through the period gap, and appear as polars
\citep[e.g.][]{hellier01-2, cumming02-1}. IPs with low field strengths
should remain unsynchronised below the period gap. This general
hypothesis has been backed by the detailed simulations of
\citet{nortonetal04-1}, who find that long-period IPs with a white
dwarf magnetic moment of
$\mu_{\rm{wd}}\gtrsim5\times10^{33}$\,G\,$\rm{cm^{3}}$ will evolve
into polars while those with
$\mu_{\rm{wd}}\lesssim5\times10^{33}$\,G\,$\rm{cm^{3}}$ and secondary
stars with weak magnetic fields will remain IPs. Historically, the
dearth of known IPs below the period gap has raised some concerns
regarding the evolution of low-field IPs, however, a number of such
systems have been identified (see e.g. \citealt{rodriguez-giletal04-1,
  pattersonetal04-1, southworthetal07-1}), suggesting that their
number has been underestimated.

We are currently investigating the population of CVs within the
galactic plane, making use of the Isaac Newton Telescope (INT)/Wide
Field Camera (WFC) Photometric $\rm{H}\alpha$ Survey of the northern
galactic plane (IPHAS, \citealt{drewetal05-1,
  gonzales-solaresetal08-1}). \citet{withametal07-1} presented the
first eleven new CVs identified within IPHAS because of their
H$\alpha$ emission. Here, we present follow-up time-resolved
photometry of the eclipsing CV, IPHAS\,J062746.41+014811.3 (hereafter
IPHAS\,J0627), suggested by \citet{withametal07-1} to be a long-period
system, and classify it as the fourth deeply eclipsing IP,
making it a promising candidate for accurate stellar parameter
  measurements. Following the determination of the orbital and spin
periods of IPHAS\,J0627, along with estimates of its binary
inclination and mass ratio, we discuss the sample of confirmed IPs
as well as the future evolution of IPHAS\,J0627.

\section{Observations and data reduction}

\subsection{Time-series photometry}

\begin{figure}
\includegraphics[width=6.5cm]{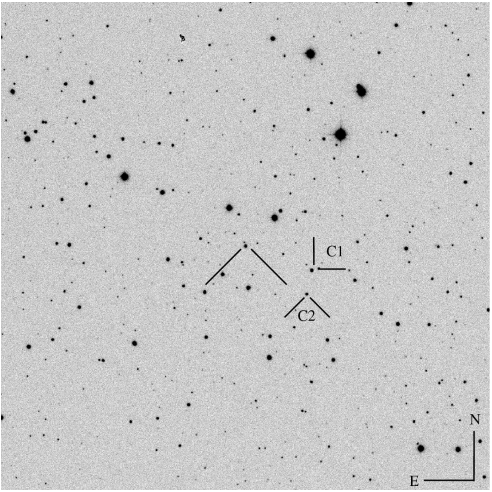}
\caption{\label{f-fc}A $7\arcmin\times7\arcmin$ finding chart of
  IPHAS\,J0627 obtained from IPHAS imaging data. The J2000 coordinates
  of the star are $\alpha=06^\mathrm{h}27^\mathrm{m}46.4^\mathrm{s}$
  and $\delta=+01\degr48\arcmin11.1\arcsec$. The comparison and check
  stars used in the photometry are marked by `C1' and 'C2',
  respectively.}
\end{figure}

\begin{figure}
\includegraphics[angle=-90,width=\columnwidth]{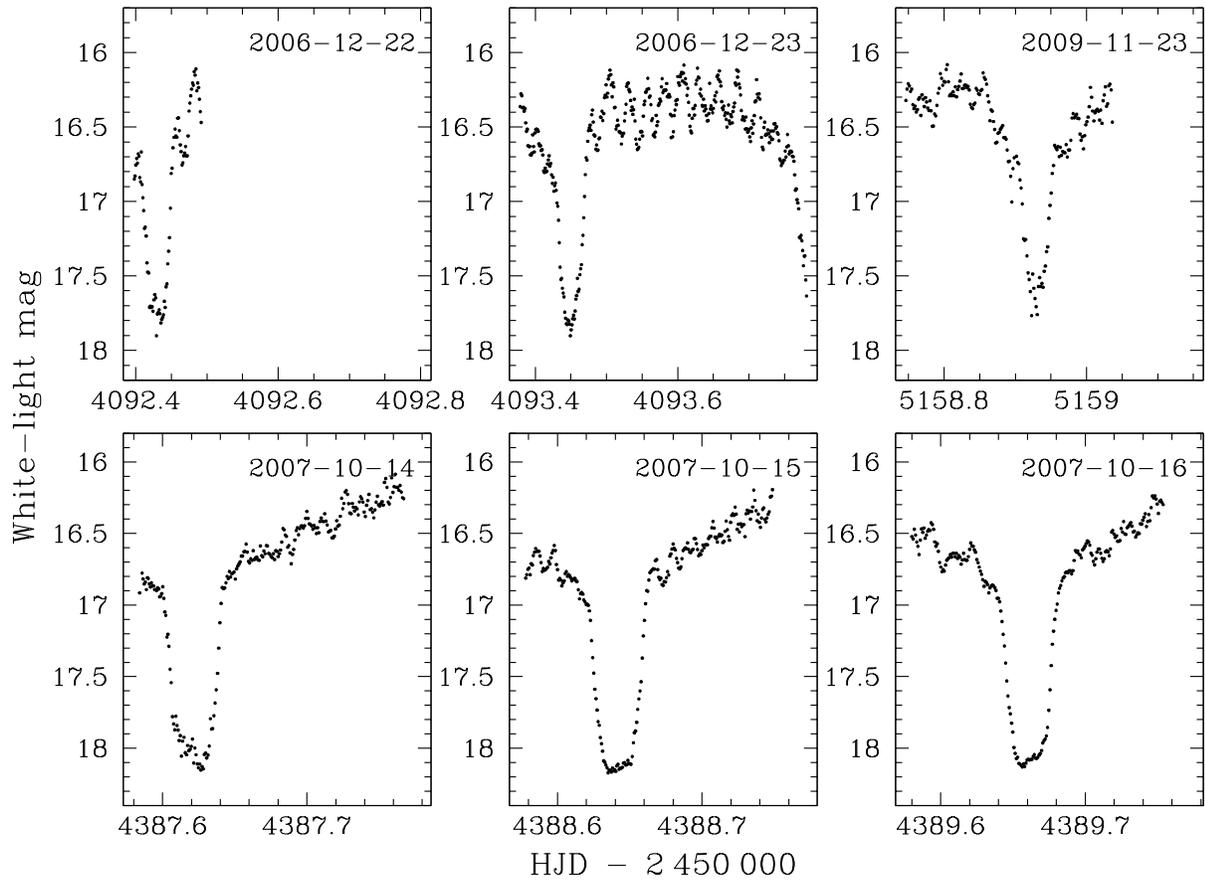}
\caption{\label{f-lc} {\it Top:} 2006 and 2009 light curves of
  IPHAS\,J0627 show similar eclipse profile. {\it Bottom:} 2007 light
  curves reveal nightly variation in the eclipse profile.}
\end{figure}

We obtained a total of $\sim 27$\,h of unfiltered time-series CCD
differential photometry of IPHAS\,J0627 (Fig.\,\ref{f-fc}) during the
period December 2006 and October 2007 at the Roque de los Muchachos
Observatory on La Palma using the 1.2\,m Mercator telescope equipped
with the $2\mathrm{k}\times2\mathrm{k}$ pixel MEROPE CCD camera
(Table\,\ref{t-obslog}). The images were taken using $3\times3$
binning to reduce the read-out noise and to improve the time
resolution. The data were reduced using the pipeline described by
\citet{gaensickeetal04-1} which employs \texttt{MIDAS} for bias
subtraction and flat fielding, and performs aperture photometry using
\texttt{Sextractor} \citep{bertin+arnouts96-1}. Differential
magnitudes of IPHAS\,J0627 were then calculated relative to the
comparison star C1 (USNO-A2.0\,0900-02977965: $R$=16.1, $B$=18.0). C2
(USNO-A2.0\,0900-02978083: $R$=17.1, $B$=18.1) was used to check for
variability of C1 which no significant brightness changes were
found. Sample light curves of IPHAS\,J0627 are shown in
Fig.\,\ref{f-lc}.

One additional light curve of IPHAS\,J0627 was obtained
quasi-simultaneous with the \textit{Swift} X-ray observations (see
below) using the AAVSOnet telescope Wright28, a C-11 equipped with an
ST-7 camera. The data were reduced in a standard fashion using
MaximDL/CCD.

\subsection{\textit{Swift} X-ray and ultraviolet data}
IPHAS\,J0627 was observed with the narrow-field instruments of the {\it
  Swift} spacecraft \citep{gehrelsetal04-1} for a total of 9\,ks on 23
November 2009. The observation was broken across nine spacecraft
orbits, with exposure times ranging from 0.2 to 1.5\,ks.

Observations with the Ultraviolet/Optical Telescope
\citep[UVOT;][]{romingetal05-1} were made using the UVM2 filter, which has
central wavelength of 217\,nm and a full-width at half-maximum bandwidth
of 51\,nm.  One exposure was made each visit. A source was visible at
the position of IPHAS\,J0627 in all nine images, and a light curve
was extracted from a 5 arcsec radius region using the {\sc
  uvotmaghist} tool version 1.12 and photometric calibration data from
the release of 22 May 2007 (version 105).

Observations with the X-ray Telescope \citep[XRT;][]{burrowsetal05-1} were made 
predominantly in photon counting mode (PC) and we did not attempt to 
analyse the 10 per cent of data collected in Windowed Timing mode (WT). 
A light curve and spectrum were extracted within a 20\,pixel 
(47\,arcsec) radius circle of the source position from the cleaned 
event file using {\sc xselect} version 2.4 and retaining events with grades 
0--12. The background was estimated using a circular region of 4.6\arcmin\
radius. The spectrum was binned to a minimum of five counts per bin. 

\section{Light curve analysis}\label{s-analysis}
The light curves in Fig.\,\ref{f-lc} confirm the deeply eclipsing
nature of IPHAS\,J0627 found by \citet{withametal07-1}. In addition,
the 2006 data exhibit two additional features: short-period
  modulation and a broad modulation of the out-of-eclipse brightness
of the system. Below, we analyse these three morphological light curve
structures.

\subsection{Eclipse profiles and ephemeris} 

\citet{withametal07-1} used two accurate eclipse times plus a rough
estimate of a third eclipse time to determine a set of four possible
orbital periods for IPHAS\,J0627, $\sim1.02$\,d, $\sim0.51$\,d,
$\sim0.34$\,d and $\sim0.25$\,d. One aim of the observations discussed
here was to measure the actual orbital period of IPHAS\,J0627 and to
determine an accurate eclipse ephemeris. For that purpose, we
determined mid-eclipse times by mirroring and shifting the eclipse
profiles until the best match in overall shape was achieved. Combining
these six new eclipse times (Table\,\ref{t-eclipse_minima}) with
those from \citet{withametal07-1}, we determined a unique cycle count
and a best-fit linear ephemeris
\begin{equation}
\label{e-ephemeris}
\T=\mathrm{HJD}\,2453340.50732(40) + 0.34008253(14)\times E
\end{equation}
where \T\ is defined as the time of mid-eclipse and the errors are
given in brackets. We hence conclude that the orbital period of
IPHAS\,J0627 is $\Porb=8.1619807(34)$\,h. The corresponding cycle
numbers and observed minus computed ($\mathrm{O-C}$) eclipse times are
reported in Table\,\ref{t-eclipse_minima}.

\begin{figure}
\includegraphics[angle=-90,width=10cm]{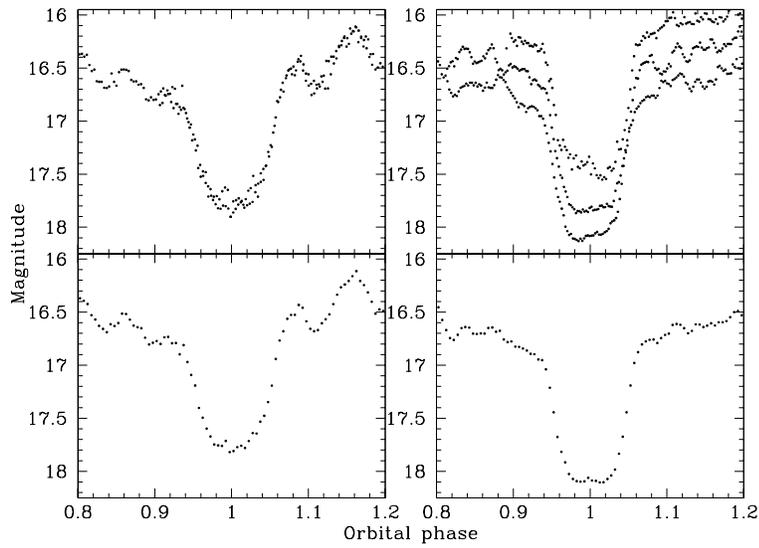}
\caption{\label{f-folded_ecl} \textit{Top:} the 2006 (left) and 2007
  (right) eclipse profiles of IPHAS\,J0627 folded on the ephemeris in
  Eq.\,\ref{e-ephemeris}. The two eclipse observations from 2006 align
  very well in shape and depth. The 2007 October 14, 15, and 16
  eclipses have been shifted by 0, -0.3, and -0.6 magnitudes to
  highlight the night-to-night variations in the eclipse profile.
  \textit{Bottom:} the same data as in the top panels, but
  averaged into phase bins of $\Delta\phi=0.005$.}
\end{figure}

The Mercator light curves folded on the ephemeris in
Eq.\,(\ref{e-ephemeris}) are shown in Fig.\,\ref{f-folded_ecl},
illustrating a noticeable change in the shape of the eclipse
profiles. The two light curves obtained on 2006 December 22 \& 23 show
nearly perfect agreement, with a relatively round-shaped bottom of the
eclipse profile, whereas the 2007 observations exhibit nightly
variation in the eclipse profile, and are overall more box-shaped.  In
2006, the eclipse depth of the average light curve was
$\simeq1.3\pm0.1$\,mag, and the full-width of the eclipse at half
depth was $\fullwidth\simeq0.115\pm0.006$ (see Sect.\,\ref{s-inc} for
details of estimating \fullwidth). In 2007, the eclipse depth was
$\sim1.43\pm0.05$\,mag, with a full-width at half depth of
$\fullwidth\simeq0.106\pm0.002$. The 2009 observations were taken with
very long exposure times, but at face value, the eclipse had a similar
round-shaped profile as in 2006. 

In addition to the change in the eclipse profile morphology,
we investigated the out-of-eclipse brightness variations by measuring
the average magnitude in the phase interval 0.8--0.9
and 1.1--1.2. These measurements suggest that the out-of-eclipse
magnitude of IPHAS\,J0627 is varying by $\sim0.2$\,mag, with the
system having been found at $\simeq16.3$\,mag, $\simeq16.5$\,mag, and
$\simeq16.3$\,mag in 2006, 2007, and 2009, respectively. The
decreased brightness level and the narrower eclipse width observed in
2007 imply that the accretion disk contributed less to the optical
light during that epoch. The flat bottom of the eclipse profile is
suggestive that the white dwarf and the accretion disk may have been
totally eclipsed, a higher time resolution study could potentially
resolve the white dwarf ingress and egress.

\subsection{Spin modulation}
In addition to the deep eclipses, the December 2006 light curves of
IPHAS\,J0627 exhibit short-period modulation on time-scales
of $\sim40$\,min with a $\sim0.4-0.5$\,mag peak-to-peak amplitude,
most clearly seen in the December 23 observations covering more than
one orbital cycle (see Fig.\,\ref{f-lc}). Considering the detection of
\Line{He}{II}{4686} emission in the spectrum of IPHAS\,J0627
\citep{withametal07-1}, this raises the possibility that the observed
oscillations represent the white dwarf spin period.

In order to test the periodicity of the oscillations, we subjected the
combined light curves 2006 December 22 and 23 observations to a
time-series analysis using the \texttt{MIDAS/TSA} context. Prior to
the analysis, the mean was subtracted from the data. In addition, we
pre-whitened the data by means of a sine fit, fixing the period of the
sine wave to the orbital. We included nine harmonic frequencies in the
sine fit to remove the effect of the eclipse from the observed light curve.

\begin{figure}
\includegraphics[width=12cm]{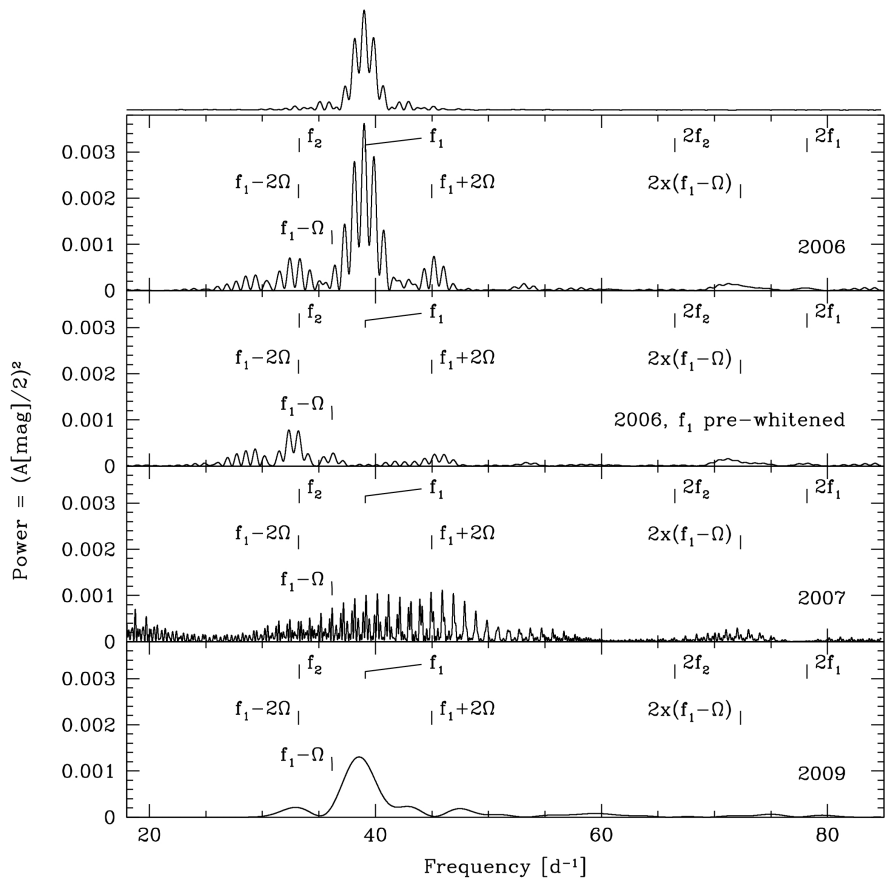}
\caption{\label{f-powerspectrum} Power spectrum computed from the
  combined photometric data obtained in December 2006 (top panel), and
  the corresponding window function (above the figure). The power
  spectrum computed from the 2006 data after pre-whitening with the
  strongest signal, $f_1=39.090(15)$\,\id, contains residual power at
  $33.244(29)$\,\id, which, within the uncertainties is consistent
  with $f_1-2\Omega$ (second panel). Additionally, there is some evidence for
  low-amplitude power near $f_1+2\Omega$ and $2(f_1-\Omega)$, whereas
  no signal is detected near the second harmonic of either $f_1$ or
  $f_2$. The power spectra from October 2007 and November
  2009 are shown in the third panel from the top, and the bottom
  panel, respectively.}
\end{figure}

\begin{figure}
\includegraphics[width=12cm]{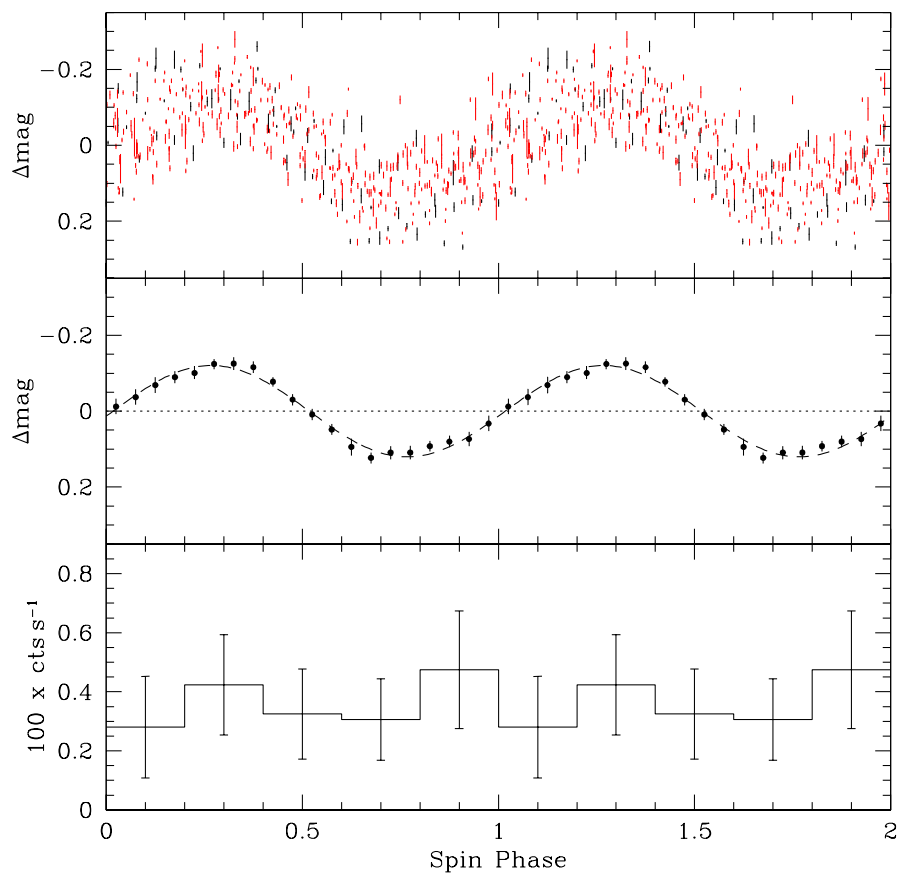}
\caption{\label{f-folded_spin} Spin-folded optical and X-ray light
  curve of IPHAS\,J0627 adopting $\Pspin=2210$\,s. The zero-point of
  the spin phase is arbitrary.} {\it Top:} all individual data points
from December 2006 (black: December 22nd, red: December
23rd). \textit{Middle:} the 2006 data binned into 20 phase slots,
along with a sine fit to the binned and folded data (dashed
line). \textit{Bottom:} \textit{Swift} XRT X-ray light curve of
IPHAS\,J0627 folded on the spin period of 2210\,s.
\end{figure}

The power spectrum computed from the data prepared in this way
contains the strongest signal at $f_1=39.090(15)$\,\id\
(Fig.\,\ref{f-powerspectrum}), flanked by one-day aliases. The
best-fit value of the period determined from a sine fit to the data is
2210.27(87)\,s.  We assessed the likelihood of correct alias choice
using a test based on bootstrapping simulations as described in
\citet{southworthetal06-1, southworthetal07-2}, and find that 100\% of
the simulations return the strongest power within the 39.090\,\id\
alias. We tested the significance of this signal by creating a faked
data set computed from a sine function with a frequency of
39.090\,\id, and randomly offset from the computed sine wave using the
observed errors. The power spectrum of the faked data set reproduces
well the 1-day alias structure of the power spectrum calculated from
the observations of IPHAS\,J0627 (Fig.\,\ref{f-powerspectrum}, {\it
top curve}). The photometric data folded on $2210$\,s display a
quasi-sinusoidal modulation with an amplitude of $\sim0.2$\,mag
(Fig.\,\ref{f-folded_spin}). Such coherent and large-amplitude optical
modulation is a hallmark of intermediate polars, e.g. FO\,Aqr
\citep{pattersonetal98-3}, AO\,Psc \citep{pattersonetal81-2}, or
MU\,Cam \citep{araujo-betancoretal03-2}. Typically, the power spectra
of IPs show signals at the orbital frequency, $\Omega$, the white
dwarf spin frequency, $\omega$, and the orbital side-bands
$\omega\pm\Omega$ and $\omega-2\Omega$
\citep[e.g.][]{warner86-2}. Inspecting the power spectrum in
the top panel of Fig.\,\ref{f-powerspectrum} reveals power in excess
of the alias structure. The strongest signal in the power spectrum
computed from the data pre-whitened with $f_1=39.090$\,\id\
(Fig.\,\ref{f-powerspectrum}, second panel from top) is found at
$f_2=33.244(29)$\,\id\, which is, within the uncertainties, equal to
$f_1-2\Omega$. Additional low-amplitude signals are seen near
$f_1+2\Omega$ and possibly $2(f_1-\Omega)$, however, longer
time-series photometry will be necessary to confirm the presence of
these signals. Based on the most commonly observed behaviour among the
known IPs, we identify the strongest signal as the white dwarf spin
frequency, $\omega=f_1$, and the weaker signal as an orbital side-band
$\omega-2\Omega$. Alternatively, $f_2$ is the spin frequency, in which
case the strongest signal would be the $\omega+2\Omega$ side-band,
however, we consider this option less likely. We hence conclude that
IPHAS\,J0627 is an eclipsing intermediate polar, and the white dwarf
spin period is most likely $\Pspin=2210.27(87)$\,s, where the error
was determined by means of a sine fit to the spin light curve.

The amplitude of the optical spin modulation undergoes large long-term
variations, as it was very weak in our short observations in October
2007 (see Fig.\,\ref{f-powerspectrum}, third panel from top). The
weakening of the spin signal in 2007 may have been caused by a lower
accretion rate, as suggested by the fainter magnitude compared to the
2006 observations. In 2009, when the system was again brighter, the
spin modulation was back, though with a lower amplitude compared to
2006 (Fig.\,\ref{f-powerspectrum}, bottom panel). The spin period
determined from that single night was found to be
$2237(10)$\,sec. Pre-whitening the light curve with a multi-harmonic
sine-fit to remove the effect of the eclipse introduces a systematic
uncertainty into measurement of the spin period, and we conclude that
the 2006 and 2009 values of the spin period are broadly consistent
with each other.

\subsection{Orbital modulation: a reflection effect?}
\begin{figure}
\includegraphics[angle=-90,width=8cm]{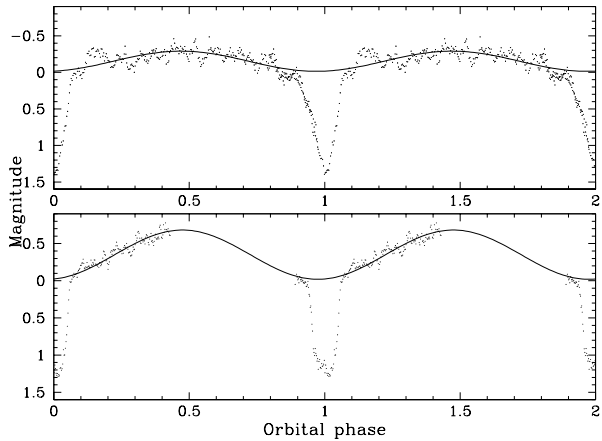}
\caption{\label{f-reflect} The orbital phase-folded light curves of
  IPHAS\,J0627 show a broad modulation, after pre-whitening
  with the adopted spin period of 2210\,s for the 2006 data ({\it top panel}),
  and raw light curve for the 2007 data ({\it bottom panel}). Fitting
  this modulation with a sine results in amplitudes of the modulation
  of $\sim0.15$\,mag in 2006 and $\sim0.33\,mag$
  in 2007.}
\end{figure}

Another distinct feature found in the light curves of IPHAS\,J0627 is
a broad modulation outside the eclipses, detected in the long
observation on 2006 December 23. This modulation may be caused by a
reflection effect, i.e. heating of the inner hemisphere of the donor
star by the accreting white dwarf, such as observed in CVs
(e.g. DD\,Cir; \citealt{woudt+warner03-1}) or in pre-CVs containing
hot primary stars (e.g. HW\,Vir; \citealt{hilditchetal96-1}, or
HS1857+5144; \citealt{aungwerojwitetal07-1}). We investigated this
modulation by pre-whitening the 2006 December 23 with the spin period,
$\Pspin=2210$\,s, and folding the data over the orbital period,
$\Porb=8.16$\,h. The 2007 October 14--16 light curves are combined and
folded on the orbital period. Phase-folded light curves are shown in
Fig.\,\ref{f-reflect} with a maximum brightness at $\phi\simeq0.5$
which is in agreement with maximum light at superior conjunction of
the secondary star when taken reflection effect into account. Fitting
a sine wave to the modulation outside the eclipse, we find the
amplitude of the modulation to be $\sim0.14$\,mag and $\sim0.33$\,mag
for the 2006 and 2007 light curves, respectively. Based on our limited
data, we suggest that the larger amplitude of the modulation observed
in 2007 may related with the fainter accretion disk contributing
somewhat less to the optical light. In order to confirm our
hypothesis, long-term observations covering the entire orbital period
are strongly encouraged.

\section{Orbital inclination}\label{s-inc}
Considering the geometry of a point eclipse by a spherical body, we
estimated the inclination, $i$, of a binary system through the
relation
\begin{equation}
\label{e-ewidth}
\bigg(\frac{R_{2}}{a}\bigg)^2 = \sin^2(\pi\fullwidth)+\cos^2(\pi\fullwidth)\cos^2 i,
\end{equation}

where $R_2/a$ is the volume radius of the secondary star, which
depends only on the mass ratio, $q=M_2/M_1$ \citep{eggleton83-1}:
\begin{equation}
\label{e-rocheradius}
\bigg(\frac{R_{2}}{a}\bigg) = \frac{0.49q^{2/3}}{0.6q^{2/3}+\ln(1+q^{1/3})}
\end{equation}

and \fullwidth\ is the full-width of eclipse at half depth (see also
e.g. \citealt{dhillonetal91-1, rodriguez-giletal04-2}). We estimated
\fullwidth\ for IPHAS\,J0627 from the 2006, 2007, and 2009 combined
light curves with an average out-of-eclipse magnitude of $16.5\pm0.1$,
$16.7\pm0.1$, and $16.6\pm0.1$, respectively. This yields
$\fullwidth\simeq0.115\pm0.006$, $\fullwidth\simeq0.106\pm0.002$, and
$\fullwidth\simeq0.120\pm0.005$ for 2006, 2007, and 2009 observations,
respectively; the large error is due to the large uncertainty in
identifying the out-of-eclipse brightness.

In order to obtain the inclination of the system, a given value of the
mass ratio, $q$, need to be assumed. Using the mean empirical
mass-period relation of \citet{smith+dhillon98-1}, 
\begin{equation}
\label{e-mass-period}
\frac{M_2}{\Msun} = (0.126\pm0.011)\Porb-(0.11\pm0.04)
\end{equation}
where \Porb\ is expressed in hours, we find $0.87\,\Msun \lesssim
\Msec \lesssim 0.97\,\Msun$ for the secondary star in
IPHAS\,J0627. \citet{ramsay00-1} estimated a mean value of
$\Mwd=0.85\pm0.21$\,\Msun\ for the white dwarf mass in intermediate
polars, which is broadly consistent with the mean white dwarf mass
across all CVs \citep{knigge06-1, littlefairetal08-1, kniggeetal11-1,
  zorotovicetal11-1}. Assuming stable mass transfer, we adopt
$0.85\,\Msun \lesssim \Mwd \lesssim 1.06\,\Msun$, resulting in $0.8
\lesssim q \lesssim 1.0$. This finally leads to an orbital inclination
of $77\degr \lesssim i \lesssim 84\degr$ which is in a good agreement
with the values derived in term of graphical form of the relationship
between \fullwidth, $i$, and $q$ for Roche geometry in
\citet{horne85-1}.

\section{The \textit{Swift} observations}
\begin{figure}
\includegraphics[width=8.4cm]{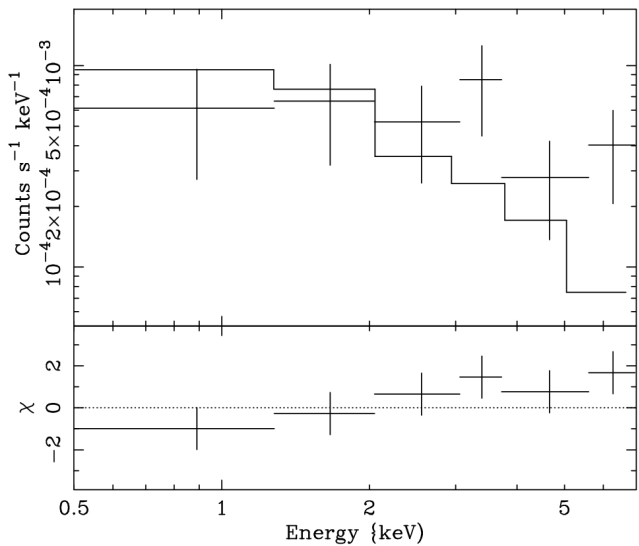}
\caption{{\it Swift} XRT X-ray spectrum of IPHAS\,J0627. The model
  curve is an optically-thin thermal plasma model with temperature of
  80\,keV. The observed spectrum is harder than this model, indicating
  the presence of absorption that is well fit by a partial-covering
  absorber (see text). }
\label{fig-xspec}
\end{figure}

A faint X-ray source was detected at the position of IPHAS\,J0627 with
a count rate of $3.2\pm0.7\rm\, ks^{-1}$. The X-ray spectrum is
plotted in Fig.\,\ref{fig-xspec} compared with the best-fitting
optically-thin thermal plasma model \citep{mewe86etal-1,
  liedahl95etal-1}.  In this fit, the temperature has risen to the
model maximum of 80\,keV, and it is clear that the observed spectrum
is harder still. The fit is only marginally acceptable with a reduced
$\chi^2$ of 1.75 with 4 degrees of freedom. Adding a cold absorber to
the model improves the fit to a reduced $\chi^2$ of 1.30 (3 degrees of
freedom) with a best-fitting $N_{\rm H}$ of
$5\times10^{21}\rm\,cm^{-2}$. The hard spectrum and high absorption
are as expected for an intermediate polar, but since the source is
located close to the Galactic Plane it is not clear whether this
absorption is intrinsic or interstellar. The total Galactic column in
the direction of IPHAS\,J0627 is also
$5\times10^{21}\rm\,cm^{-2}$. However, the fit is further improved by
allowing the absorber to only partially cover the source, with a
higher column density of $N_{\rm H}=4\times10^{22}\rm\,cm^{-2}$, a
partial-covering fraction of 0.9, and a temperature that is no longer
forced the highest allowed values, $kT=5$\,keV. This fit yields a
reduced $\chi^2$ of 0.96 with 2 degrees of freedom. Although the
signal to noise ratio is low, we can conclude that the X-ray spectrum
of IPHAS\,J0627 is consistent with that expected for an intermediate
polar. The 0.5--10\,keV flux of the best-fitting model is
$2.2\times10^{-13}\,\rm erg\,s^{-1}\,cm^{-2}$. This is about an order
of magnitude fainter than would be expected from scaling by the
optical fluxes of well studied (and usually X-ray selected) IPs
\citep[e.g.][]{landietal09-1, brunschweigeretal09-1,
  scaringietal10-1}.

In order to search for the presence of a white-dwarf spin modulation
in the X-ray data we folded the XRT light on the period of 2210\,s.
The folded light curve is presented in Fig.\,\ref{f-folded_spin}
(bottom panel) and does not show any sign of a modulation at this
period. However, with such a low number of events detected, the 90 per
cent confidence upper limit on the amplitude of a sinusoidal
modulation is 65 per cent. So the \textit{Swift} data do not rule out
the presence of an X-ray spin modulation in this object.  A Fourier
analysis of the X-ray light curve also failed to reveal any other
significant periods.

The \textit{Swift} ultraviolet data were obtained in the imaging mode,
i.e. no time information is available for individual photons, but only
average ultraviolet fluxes for each of the nine spacecraft orbits. The
one ultraviolet measurement made close to the optical eclipse phase
also has the lowest flux, indicating that the eclipse is also present
at ultraviolet wavelengths. Excluding the eclipse, the ultraviolet
flux at 217\,nm varies in the range 5--13$\times10^{-17}\,\rm
erg\,s^{-1}\,cm^{-2}$, exceeding the statistical errors on the flux
individual measurements.

\section{Discussion}

Over the past few years, the number of confirmed intermediate polars
has rapidly increased. At the time of writing, the IP page by
K. Mukai\footnote{http://asd.gsfc.nasa.gov/Koji.Mukai/iphome/iphome.html}
lists 36 confirmed IPs while \citet[][v.7.12]{ritter+kolb03-1}
contains roughly twice this number, which underlines the rather broad
range of criteria adopted by different authors to classify a system as
IP. One clear hallmark of IPs is the presence of coherent optical
and/or X-ray short-term variability on the white dwarf spin period
over a sufficient span of time \citep[e.g.][]{buckley00-1}.

Detailed measurements of the physical parameters of CVs come from
observational studies of eclipsing systems. Mukai's IP list contains
only six confirmed eclipsing IPs, of which four only show
grazing/partial eclipses: FO\,Aqr \citep[e.g.][]{hellieretal90-1,
kruszewski+semeniuk93-1}, BG\,CMi \citep[e.g.][]{patterson+thomas93-1,
kimetal05-1}, TV\,Col \citep[e.g.][]{hellieretal91-1, hellier93-1},
EX\,Hya \citep[e.g.][]{beuermann+osborne88-1}. The other two, DQ\,Her
\citep[e.g.][]{walker54-1, walker56-1} and XY\,Ari
\citep{patterson+halpern90-1} are deeply eclipsing IPs.  Detailed
observational and theoretical studies of DQ\,Her provided tight
constraints on its system parameters, i.e. $\Porb$, $\fullwidth$, $q$,
$i$, $\Mwd$, $\Msec$, and disk radius \citep[see
e.g.][]{horneetal93-1, zhangetal95-1}. XY\,Ari exhibits deep X-ray
eclipses, but is hidden behind the molecular cloud MBM12 which makes
it virtually invisible in the optical band \citep{littlefairetal01-1}.
Recently, \citet{warner+woudt09-1} identified V597\,Pup as a third
deeply eclipsing ($\simeq1$\,mag depth) IP, which is in the stage of
decline to its pre-eruption brightness at $V\sim20$.

Based on the optical short-period variation at $\Pspin=2210$\,s
detected in our 2006 light curves, we classify IPHAS\,J0627 as the
fourth deeply eclipsing IP with $\Porb=8.16$\,h, turning it to a rare
object that holds substantial promises for detailed optical and X-ray
follow-up studies.

\begin{figure}
\includegraphics[width=8cm]{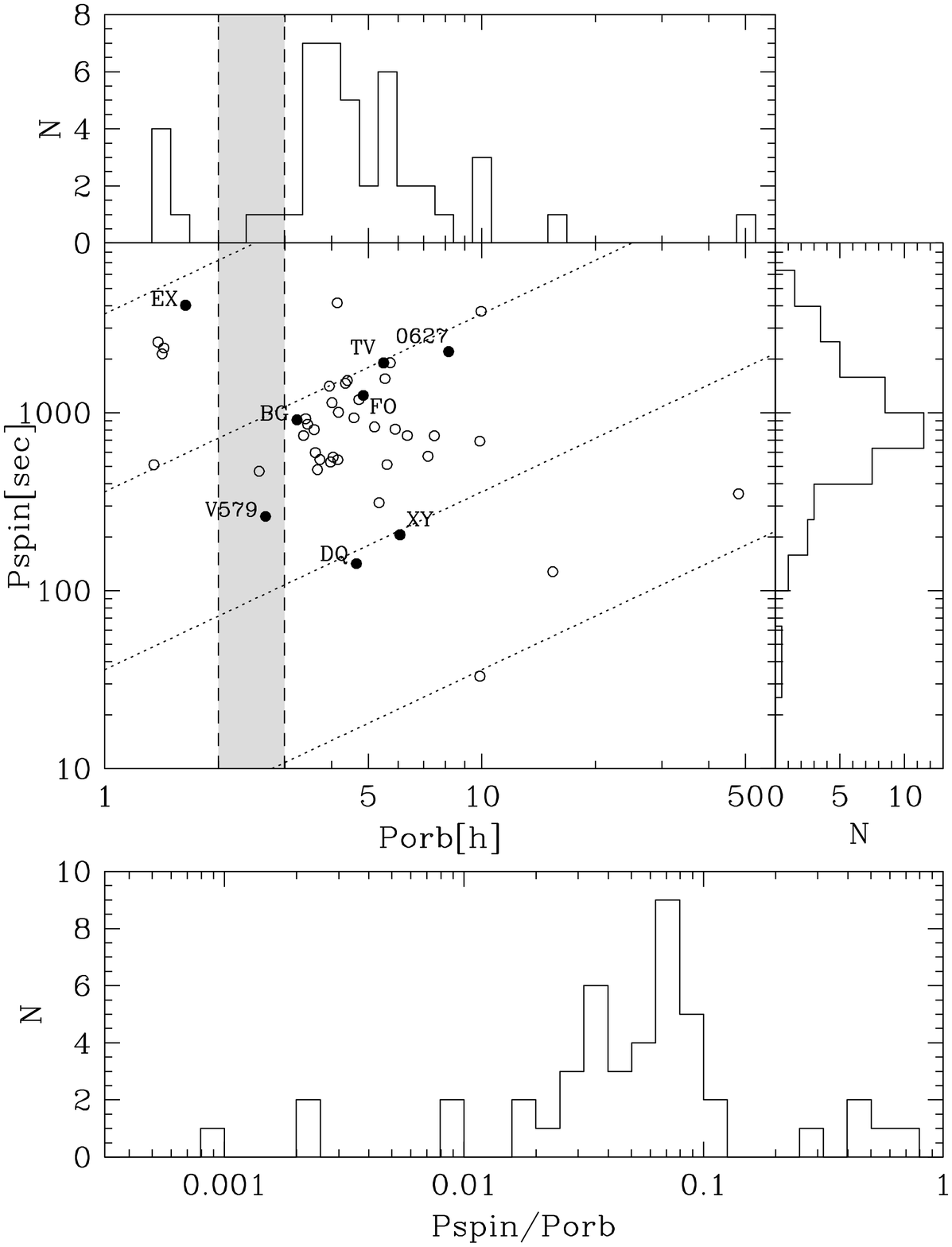}
\caption{\label{f-ip} The updated period distribution of IPs on the
original of \citet{gaensickeetal05-1}. {\it Middle panel:} orbital and
spin period of 48 IPs. The dotted lines indicate
$\Pspin/\Porb=1,0.1,0.01,0.001$ from top to bottom, respectively. The
eclipsing systems are shown as filled symbols.} {\it Top panel:} orbital
period distribution of the known IPs, the 2--3\,h period gap is shaded
grey. {\it Right panel:} spin period distribution of the known IPs.
\end{figure}

We adopted Mukai's conservative classification, and updated his list
with additional 9 IPs: V597\,Pup, IGR\,J16500-3307, IGR\,J17195-4100,
IGR\,J19267+1325, 1RXS\,J165443.5-191620, IGR\,J08390-4833,
IGR\,J18308-1232, IGR\,J18173-2509, IPHAS\,J0627, and 3 IPs from
Fig.\,23 of \citet{gaensickeetal05-1} i.e., RXJ0153.3+7446,
HS\,0943+1404, 1RXS\,J063631.9+353537. Figure\,\ref{f-ip} shows the
most up-to-date distribution of the 48 confirmed IPs in the
$\Pspin-\Porb$ plane (updated with respect to Fig.\,23 of
\citet{gaensickeetal05-1} and with the additional well-determined
$\Porb$ and $\Pspin$ IPs listed in Table\,\ref{t-IPs}). Eclipsing
systems presented as filled dots. It is clear that the majority of IPs
($\sim87\%$) are found above the conventional 2--3\,h period gap
whilst the fraction of systems below the period gap remains fairly
small ($\sim13\%$). Only two systems have extremely long orbital
periods i.e., GK\,Per ($\Porb=1.996$\,d; \citealt{cramptonetal86-1})
and 1RXS\,J173021.5-055933 ($\Porb=15.42$\,h;
\citealt{gaensickeetal05-1}).

The updated distribution shows that a fair number of CVs have
$\Pspin/\Porb\simeq0.1$, a trend already noticed frequently in the
past \citep[e.g.][]{barrettetal88-1, nortonetal04-1,
gaensickeetal05-1, scaringietal10-1}, which spawned the initial
theoretical work on the white dwarf equilibrium in magnetic CVs
\citep{king+lasota91-1, warner+wickramasinghe91-1}. However, it is now
clear that IPs above the period gap (3--10\,h) are widely distributed
over $0.01\la\Pspin/\Porb\la0.1$, including IPHAS\,J0627 with
$\Pspin/\Porb=0.075$ (for the adopted $P_\mathrm{spin}=2210$\,s),
indicating disk-fed accretion \citep{nortonetal04-1,
nortonetal08-1}. All IPs with $\Porb<2$\,h have $\Pspin/\Porb>0.1$
which agrees with the predictions of \citet{king+wynn99-1}. The most
extreme systems with $\Pspin/\Porb<0.01$ are exclusively found at very
long orbital periods, which may suggest that they are relatively young
systems still far from equilibrium.

\citet{nortonetal04-1} showed that a large range of spin equilibria
exists in the $(\Pspin/\Porb, \Porb, \mu_{\rm{wd}}, q)$ parameter
plane, with $\mu_{\rm{wd}}$ being the magnetic moment of the white
dwarf, as illustrated for a mass ratio $q=0.5$ in their Fig.\,2.  For
$\Porb=8.16$\,h, $P_\mathrm{spin}/P_\mathrm{orb}=0.075$ (adopting a
spin period of 2210\,s), and correcting for the higher mass ratio of
IPHAS\,J0627 ($q\simeq0.8$, see Eq.\,11 of \citealt{nortonetal04-1}),
we estimate from the Fig.\,2
$\mu_{\rm{wd}}\sim6-7\times10^{33}$\,G\,$\rm{cm^{3}}$. With such a
relatively high magnetic moment, IPHAS\,J0627 may just about evolve
into a short-period EX\,Hya-like IP, with a large
$P_\mathrm{spin}/P_\mathrm{orb}$ ratio, or, perhaps more likely,
synchronise as a polar. In fact, adopting $R_\mathrm{wd}=0.01$\,\Rsun
(appropriate for the average CV white dwarf mass of 0.85\,\Msun), the
estimated magnetic moment implies a field strength of $B\simeq18$\,MG,
which comparable to that of the short-period polars EF\,Eri and
ST\,LMi.

\begin{figure}
\includegraphics[angle=-90, width=8.4cm]{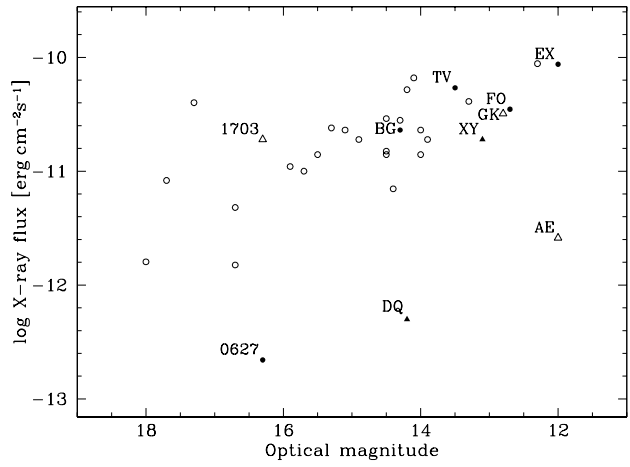}
\caption{X-ray fluxes and optical magnitudes of the confirmed
IPs. Filled dots represent eclipsing systems. Filled triangles are
eclipsing and rapid rotators. Open triangles are rapid rotators.}
\label{f-fxfopt}
\end{figure}

The motivation of our \textit{Swift} observation of IPHAS\,J0627 was
to probe for X-ray emission pulsed on the white dwarf spin period,
which would be the ultimate confirmation of the IP nature of this
system. We found that the best-fitting model at 0.5--10\,keV flux for
IPHAS\,J0625 is $2.2\times10^{-13}\,\rm erg\,s^{-1}\,cm^{-2}$. This
value is an order of magnitude fainter than most confirmed IPs which
usually are X-ray selected. Figure\,\ref{f-fxfopt} presents X-ray
fluxes and optical magnitudes of the confirmed IPs\footnote{All X-ray
fluxes used in Fig.\,\ref{f-fxfopt} were taken from Mukai's list with
2-10\,keV fluxes except DQ\,Her \citep{patterson94-1},
1\,RXSJ070407.9+262501 \citep{anzolinetal08-1}, MU\,Cam
(=\,IGR\,J06253+7334), 1RXS\,J173021.5-055933 (=\,IGR\,J17303-0601),
IGR\,J16500-3307, IGR\,J17195-4100, V2069-Cyg \citep{landietal09-1},
IGR\,J00234+6141 \citep{anzolinetal09-1}, IGR\,J08390-4833,
IGR\,J18308-1232, IGR\,J18173-2509 \citep{bernardinietal12-1}}, and
optical magnitudes were taken from
\citet[][v.7.12]{ritter+kolb03-1}, with filled dots
represented eclipsing systems, triangles being rapid rotators
($\Pspin/\Porb<0.01$), and filled triangles being eclipsing and rapid
rotators. IPHAS\,J0627 has clearly the lowest X-ray-to-optical flux
ratio, followed by DQ\,Her and AE\,Aqr. The low X-ray flux in AE\,Aqr
is explained by the very rapid rotation of the white dwarf, which
prevents accretion \citep{wynnetal97-1}. Among the other two rapid
rotators, DQ\,Her has a low X-ray flux, but 1RXS\,J173021.5-055933 is
X-ray bright~--~both have spin periods $3-4$ times longer than
AE\,Aqr, suggesting that inefficient accretion is not necessarily the
reason for the low X-ray flux of DQ\,Her. The other plausible
hypothesis is that the X-ray flux in DQ\,Her is blocked by the
accretion disk/rim because of the high binary inclination
($i=86.5\degr$, \citealt{horneetal93-1}). To complicate the matters,
XY\,Ari is a deeply eclipsing ($i<84\degr$, \citealt{hellier97-1}),
but X-ray bright IP. However, it is difficult to assess an 'intrinsic'
X-ray-flux-to-optical ratio for XY\,Ari since the system lies behind
the molecular cloud MBM12. For partial/grazing eclipsing IPs, X-ray
fluxes are typically consistent with non-eclipsing systems.

We conclude that the dependence of the X-ray-to-optical flux ratio on
the binary inclination and white dwarf spin is not straight-forward,
but for the case of IPHAS\,J0627 obscuration of the accretion spots on
the white dwarf by the accretion disk/rim appears to be the most
likely explanation for the low X-ray flux. High-speed ground-based
photometry of IPHAS\,J0627 has the potential to settle the question
whether or not the white dwarf is hidden from direct view.

\section{Conclusions}
We have identified IPHAS\,J0627.41+014811.3 as the fourth deep
eclipsing IP with an orbital period of $\Porb=8.1619807(34)$\,h, and a
spin period of $\Pspin=2210.27(87)$\,s. Because of its eclipsing
nature, this IP is particularly well suited for detailed follow-up
studies that will provide detailed and accurate insight into the
system parameters. Our photometric data spanning three observing
seasons reveal variations in the system brightness, the amplitude of
the optical spin modulation, and the morphology of the eclipse
profiles, all of which can tentatively be explained by a variation in
the accretion rate. The relatively large magnetic moment of the white
dwarf in IPHAS\,J0627 suggests that it is right at the boundary of
systems evolving into either short-period EX\,Hya IPs or synchronised
polars.

\acknowledgments

This work is supported by the Thailand Research Fund under grant
  number MRG5180136. We gratefully acknowledge the observations of
  IPHAS\,J0627 taken through AAVSOnet, operated by the American
  Association of Variable Star Observers. We thank the referee for
  his/her constructive comments which have improved the paper.

{\it Facilities:} \facility{Mercator1.2m}, \facility{Swift},
\facility{AAVSO}

\bibliographystyle{apj} 
\bibliography{aamnem99,aabib,proceedings}

\begin{thebibliography}{82}
\expandafter\ifx\csname natexlab\endcsname\relax\def\natexlab#1{#1}\fi

\bibitem[{{Anzolin} {et~al.}(2008){Anzolin}, {de Martino}, {Bonnet-Bidaud},
  {Mouchet}, {G{\"a}nsicke}, {Matt}, \& {Mukai}}]{anzolinetal08-1}
{Anzolin}, G., {de Martino}, D., {Bonnet-Bidaud}, J.-M., {Mouchet}, M.,
  {G{\"a}nsicke}, B.~T., {Matt}, G., \& {Mukai}, K. 2008, aa, 489, 1243

\bibitem[{{Anzolin} {et~al.}(2009){Anzolin}, {de Martino}, {Falanga}, {Mukai},
  {Bonnet-Bidaud}, {Mouchet}, {Terada}, \& {Ishida}}]{anzolinetal09-1}
{Anzolin}, G., {de Martino}, D., {Falanga}, M., {Mukai}, K., {Bonnet-Bidaud},
  J.-M., {Mouchet}, M., {Terada}, Y., \& {Ishida}, M. 2009, aa, 501, 1047

\bibitem[{{Araujo-Betancor} {et~al.}(2003){Araujo-Betancor}, {G{\"a}nsicke},
  {Hagen}, {Rodr\'iguez-Gil}, \& {Engels}}]{araujo-betancoretal03-2}
{Araujo-Betancor}, S., {G{\"a}nsicke}, B.~T., {Hagen}, H.-J.,
  {Rodr\'iguez-Gil}, P., \& {Engels}, D. 2003, A\&A, 406, 213

\bibitem[{{Aungwerojwit} {et~al.}(2007){Aungwerojwit}, {G{\"a}nsicke},
  {Rodr{\'{\i}}guez-Gil}, {Hagen}, {Giannakis}, {Papadimitriou}, {Allende
  Prieto}, \& {Engels}}]{aungwerojwitetal07-1}
{Aungwerojwit}, A., {G{\"a}nsicke}, B.~T., {Rodr{\'{\i}}guez-Gil}, P., {Hagen},
  H.-J., {Giannakis}, O., {Papadimitriou}, C., {Allende Prieto}, C., \&
  {Engels}, D. 2007, A\&A, 469, 297

\bibitem[{{Barrett} {et~al.}(1988){Barrett}, {O'Donoghue}, \&
  {Warner}}]{barrettetal88-1}
{Barrett}, P., {O'Donoghue}, D., \& {Warner}, B. 1988, MNRAS, 233, 759

\bibitem[{{Bernardini} {et~al.}(2012){Bernardini}, {de Martino}, {Falanga},
  {Mukai}, {Matt}, {Bonnet-Bidaud}, {Masetti}, \&
  {Mouchet}}]{bernardinietal12-1}
{Bernardini}, F., {de Martino}, D., {Falanga}, M., {Mukai}, K., {Matt}, G.,
  {Bonnet-Bidaud}, J.-M., {Masetti}, N., \& {Mouchet}, M. 2012, A\&A, 542, A22

\bibitem[{{Bertin} \& {Arnouts}(1996)}]{bertin+arnouts96-1}
{Bertin}, E. \& {Arnouts}, S. 1996, A\&AS, 117, 393

\bibitem[{{Beuermann} \& {Osborne}(1988)}]{beuermann+osborne88-1}
{Beuermann}, K. \& {Osborne}, J.~P. 1988, A\&A, 189, 128

\bibitem[{{Bonnet-Bidaud} {et~al.}(2007){Bonnet-Bidaud}, {de Martino},
  {Falanga}, {Mouchet}, \& {Masetti}}]{bonnet-bidaudetal07-1}
{Bonnet-Bidaud}, J.~M., {de Martino}, D., {Falanga}, M., {Mouchet}, M., \&
  {Masetti}, N. 2007, A\&A, 473, 185

\bibitem[{{Bonnet-Bidaud} {et~al.}(2009){Bonnet-Bidaud}, {de Martino}, \&
  {Mouchet}}]{bonnet-bidaudetal09-1}
{Bonnet-Bidaud}, J.~M., {de Martino}, D., \& {Mouchet}, M. 2009, The
  Astronomer's Telegram, 1895, 1

\bibitem[{{Bonnet-Bidaud} {et~al.}(2006){Bonnet-Bidaud}, {Mouchet}, {de
  Martino}, {Silvotti}, \& {Motch}}]{bonnet-bidaudetal06-1}
{Bonnet-Bidaud}, J.~M., {Mouchet}, M., {de Martino}, D., {Silvotti}, R., \&
  {Motch}, C. 2006, A\&A, 445, 1037

\bibitem[{{Brunschweiger} {et~al.}(2009){Brunschweiger}, {Greiner}, {Ajello},
  \& {Osborne}}]{brunschweigeretal09-1}
{Brunschweiger}, J., {Greiner}, J., {Ajello}, M., \& {Osborne}, J. 2009, aa,
  496, 121

\bibitem[{{Buckley}(2000)}]{buckley00-1}
{Buckley}, D.~A.~H. 2000, New Astronomy, 44, 63

\bibitem[{{Burrows} {et~al.}(2005){Burrows}, {Hill}, {Nousek}, {Kennea},
  {Wells}, {Osborne}, {Abbey}, {Beardmore}, {Mukerjee}, {Short}, {Chincarini},
  {Campana}, {Citterio}, {Moretti}, {Pagani}, {Tagliaferri}, {Giommi},
  {Capalbi}, {Tamburelli}, {Angelini}, {Cusumano}, {Br{\"a}uninger}, {Burkert},
  \& {Hartner}}]{burrowsetal05-1}
{Burrows}, D.~N., {Hill}, J.~E., {Nousek}, J.~A., {Kennea}, J.~A., {Wells}, A.,
  {Osborne}, J.~P., {Abbey}, A.~F., {Beardmore}, A., {Mukerjee}, K., {Short},
  A.~D.~T., {Chincarini}, G., {Campana}, S., {Citterio}, O., {Moretti}, A.,
  {Pagani}, C., {Tagliaferri}, G., {Giommi}, P., {Capalbi}, M., {Tamburelli},
  F., {Angelini}, L., {Cusumano}, G., {Br{\"a}uninger}, H.~W., {Burkert}, W.,
  \& {Hartner}, G.~D. 2005, Space Science Reviews, 120, 165

\bibitem[{{Butters} {et~al.}(2007){Butters}, {Barlow}, {Norton}, \&
  {Mukai}}]{buttersetal07-1}
{Butters}, O.~W., {Barlow}, E.~J., {Norton}, A.~J., \& {Mukai}, K. 2007, A\&A,
  475, L29

\bibitem[{{Butters} {et~al.}(2008){Butters}, {Norton}, {Hakala}, {Mukai}, \&
  {Barlow}}]{buttersetal08-1}
{Butters}, O.~W., {Norton}, A.~J., {Hakala}, P., {Mukai}, K., \& {Barlow},
  E.~J. 2008, A\&A, 487, 271

\bibitem[{{Crampton} {et~al.}(1986){Crampton}, {Fisher}, \&
  {Cowley}}]{cramptonetal86-1}
{Crampton}, D., {Fisher}, W.~A., \& {Cowley}, A.~P. 1986, ApJ, 300, 788

\bibitem[{{Cumming}(2002)}]{cumming02-1}
{Cumming}, A. 2002, MNRAS, 333, 589

\bibitem[{{Dhillon} {et~al.}(1991){Dhillon}, {Marsh}, \&
  {Jones}}]{dhillonetal91-1}
{Dhillon}, V.~S., {Marsh}, T.~R., \& {Jones}, D. H.~P. 1991, MNRAS, 252, 342

\bibitem[{{Drew} {et~al.}(2005){Drew}, {Greimel}, {Irwin}, {Aungwerojwit},
  {Barlow}, {Corradi}, {Drake}, {G{\"a}nsicke}, {Groot}, {Hales}, {Hopewell},
  {Irwin}, {Knigge}, {Leisy}, {Lennon}, {Mampaso}, {Masheder}, {Matsuura},
  {Morales-Rueda}, {Morris}, {Parker}, {Phillipps}, {Rodriguez-Gil}, {Roelofs},
  {Skillen}, {Sokoloski}, {Steeghs}, {Unruh}, {Viironen}, {Vink}, {Walton},
  {Witham}, {Wright}, {Zijlstra}, \& {Zurita}}]{drewetal05-1}
{Drew}, J.~E., {Greimel}, R., {Irwin}, M.~J., {Aungwerojwit}, A., {Barlow},
  M.~J., {Corradi}, R.~L.~M., {Drake}, J.~J., {G{\"a}nsicke}, B.~T., {Groot},
  P., {Hales}, A., {Hopewell}, E.~C., {Irwin}, J., {Knigge}, C., {Leisy}, P.,
  {Lennon}, D.~J., {Mampaso}, A., {Masheder}, M.~R.~W., {Matsuura}, M.,
  {Morales-Rueda}, L., {Morris}, R.~A.~H., {Parker}, Q.~A., {Phillipps}, S.,
  {Rodriguez-Gil}, P., {Roelofs}, G., {Skillen}, I., {Sokoloski}, J.~L.,
  {Steeghs}, D., {Unruh}, Y.~C., {Viironen}, K., {Vink}, J.~S., {Walton},
  N.~A., {Witham}, A., {Wright}, N., {Zijlstra}, A.~A., \& {Zurita}, A. 2005,
  MNRAS, 362, 753

\bibitem[{{Eggleton}(1983)}]{eggleton83-1}
{Eggleton}, P.~P. 1983, ApJ, 268, 368

\bibitem[{{Evans} {et~al.}(2008){Evans}, {Beardmore}, \&
  {Osborne}}]{evansetal08-1}
{Evans}, P.~A., {Beardmore}, A.~P., \& {Osborne}, J.~P. 2008, The Astronomer's
  Telegram, 1669, 1

\bibitem[{{G{\"a}nsicke} {et~al.}(2004){G{\"a}nsicke}, {Araujo-Betancor},
  {Hagen}, {Harlaftis}, {Kitsionas}, {Dreizler}, \&
  {Engels}}]{gaensickeetal04-1}
{G{\"a}nsicke}, B.~T., {Araujo-Betancor}, S., {Hagen}, H.-J., {Harlaftis},
  E.~T., {Kitsionas}, S., {Dreizler}, S., \& {Engels}, D. 2004, A\&A, 418, 265

\bibitem[{{G{\"a}nsicke} {et~al.}(2005){G{\"a}nsicke}, {Marsh}, {Edge},
  {Rodr{\'{\i}}guez-Gil}, {Steeghs}, {Araujo-Betancor}, {Harlaftis},
  {Giannakis}, {Pyrzas}, {Morales-Rueda}, \&
  {Aungwerojwit}}]{gaensickeetal05-1}
{G{\"a}nsicke}, B.~T., {Marsh}, T.~R., {Edge}, A., {Rodr{\'{\i}}guez-Gil}, P.,
  {Steeghs}, D., {Araujo-Betancor}, S., {Harlaftis}, E., {Giannakis}, O.,
  {Pyrzas}, S., {Morales-Rueda}, L., \& {Aungwerojwit}, A. 2005, MNRAS, 361,
  141

\bibitem[{{Gehrels} {et~al.}(2004){Gehrels}, {Chincarini}, {Giommi}, {Mason},
  {Nousek}, {Wells}, {White}, {Barthelmy}, {Burrows}, {Cominsky}, {Hurley},
  {Marshall}, {M{\'e}sz{\'a}ros}, {Roming}, {Angelini}, {Barbier}, {Belloni},
  {Campana}, {Caraveo}, {Chester}, {Citterio}, {Cline}, {Cropper}, {Cummings},
  {Dean}, {Feigelson}, {Fenimore}, {Frail}, {Fruchter}, {Garmire}, {Gendreau},
  {Ghisellini}, {Greiner}, {Hill}, {Hunsberger}, {Krimm}, {Kulkarni}, {Kumar},
  {Lebrun}, {Lloyd-Ronning}, {Markwardt}, {Mattson}, {Mushotzky}, {Norris},
  {Osborne}, {Paczynski}, {Palmer}, {Park}, {Parsons}, {Paul}, {Rees},
  {Reynolds}, {Rhoads}, {Sasseen}, {Schaefer}, {Short}, {Smale}, {Smith},
  {Stella}, {Tagliaferri}, {Takahashi}, {Tashiro}, {Townsley}, {Tueller},
  {Turner}, {Vietri}, {Voges}, {Ward}, {Willingale}, {Zerbi}, \&
  {Zhang}}]{gehrelsetal04-1}
{Gehrels}, N., {Chincarini}, G., {Giommi}, P., {Mason}, K.~O., {Nousek}, J.~A.,
  {Wells}, A.~A., {White}, N.~E., {Barthelmy}, S.~D., {Burrows}, D.~N.,
  {Cominsky}, L.~R., {Hurley}, K.~C., {Marshall}, F.~E., {M{\'e}sz{\'a}ros},
  P., {Roming}, P.~W.~A., {Angelini}, L., {Barbier}, L.~M., {Belloni}, T.,
  {Campana}, S., {Caraveo}, P.~A., {Chester}, M.~M., {Citterio}, O., {Cline},
  T.~L., {Cropper}, M.~S., {Cummings}, J.~R., {Dean}, A.~J., {Feigelson},
  E.~D., {Fenimore}, E.~E., {Frail}, D.~A., {Fruchter}, A.~S., {Garmire},
  G.~P., {Gendreau}, K., {Ghisellini}, G., {Greiner}, J., {Hill}, J.~E.,
  {Hunsberger}, S.~D., {Krimm}, H.~A., {Kulkarni}, S.~R., {Kumar}, P.,
  {Lebrun}, F., {Lloyd-Ronning}, N.~M., {Markwardt}, C.~B., {Mattson}, B.~J.,
  {Mushotzky}, R.~F., {Norris}, J.~P., {Osborne}, J., {Paczynski}, B.,
  {Palmer}, D.~M., {Park}, H., {Parsons}, A.~M., {Paul}, J., {Rees}, M.~J.,
  {Reynolds}, C.~S., {Rhoads}, J.~E., {Sasseen}, T.~P., {Schaefer}, B.~E.,
  {Short}, A.~T., {Smale}, A.~P., {Smith}, I.~A., {Stella}, L., {Tagliaferri},
  G., {Takahashi}, T., {Tashiro}, M., {Townsley}, L.~K., {Tueller}, J.,
  {Turner}, M.~J.~L., {Vietri}, M., {Voges}, W., {Ward}, M.~J., {Willingale},
  R., {Zerbi}, F.~M., \& {Zhang}, W.~W. 2004, \apj, 611, 1005

\bibitem[{{Gonz{\'a}lez-Solares} {et~al.}(2008){Gonz{\'a}lez-Solares},
  {Walton}, {Greimel}, {Drew}, {Irwin}, {Sale}, {Andrews}, {Aungwerojwit},
  {Barlow}, {van den Besselaar}, {Corradi}, {G{\"a}nsicke}, {Groot}, {Hales},
  {Hopewell}, {Hu}, {Irwin}, {Knigge}, {Lagadec}, {Leisy}, {Lewis}, {Mampaso},
  {Matsuura}, {Moont}, {Morales-Rueda}, {Morris}, {Naylor}, {Parker}, {Prema},
  {Pyrzas}, {Rixon}, {Rodr{\'{\i}}guez-Gil}, {Roelofs}, {Sabin}, {Skillen},
  {Suso}, {Tata}, {Viironen}, {Vink}, {Witham}, {Wright}, {Zijlstra}, {Zurita},
  {Drake}, {Fabregat}, {Lennon}, {Lucas}, {Mart{\'{\i}}n}, {Phillipps},
  {Steeghs}, \& {Unruh}}]{gonzales-solaresetal08-1}
{Gonz{\'a}lez-Solares}, E.~A., {Walton}, N.~A., {Greimel}, R., {Drew}, J.~E.,
  {Irwin}, M.~J., {Sale}, S.~E., {Andrews}, K., {Aungwerojwit}, A., {Barlow},
  M.~J., {van den Besselaar}, E., {Corradi}, R.~L.~M., {G{\"a}nsicke}, B.~T.,
  {Groot}, P.~J., {Hales}, A.~S., {Hopewell}, E.~C., {Hu}, H., {Irwin}, J.,
  {Knigge}, C., {Lagadec}, E., {Leisy}, P., {Lewis}, J.~R., {Mampaso}, A.,
  {Matsuura}, M., {Moont}, B., {Morales-Rueda}, L., {Morris}, R.~A.~H.,
  {Naylor}, T., {Parker}, Q.~A., {Prema}, P., {Pyrzas}, S., {Rixon}, G.~T.,
  {Rodr{\'{\i}}guez-Gil}, P., {Roelofs}, G., {Sabin}, L., {Skillen}, I.,
  {Suso}, J., {Tata}, R., {Viironen}, K., {Vink}, J.~S., {Witham}, A.,
  {Wright}, N.~J., {Zijlstra}, A.~A., {Zurita}, A., {Drake}, J., {Fabregat},
  J., {Lennon}, D.~J., {Lucas}, P.~W., {Mart{\'{\i}}n}, E.~L., {Phillipps}, S.,
  {Steeghs}, D., \& {Unruh}, Y.~C. 2008, MNRAS, 388, 89

\bibitem[{{Gotthelf} \& {Halpern}(2010)}]{gotthelf+halpern10-1}
{Gotthelf}, J.~P. \& {Halpern}, E.~V. 2010, The Astronomer's Telegram, 2681, 1

\bibitem[{{Hellier}(1993)}]{hellier93-1}
{Hellier}, C. 1993, MNRAS, 264, 132

\bibitem[{{Hellier}(1997)}]{hellier97-1}
---. 1997, MNRAS, 291, 71

\bibitem[{{Hellier}(2001)}]{hellier01-2}
---. 2001, Cataclysmic Variable Stars (Springer)

\bibitem[{{Hellier} {et~al.}(1990){Hellier}, {Mason}, \&
  {Cropper}}]{hellieretal90-1}
{Hellier}, C., {Mason}, K.~O., \& {Cropper}, M. 1990, MNRAS, 242, 250

\bibitem[{{Hellier} {et~al.}(1991){Hellier}, {Mason}, \&
  {Mittaz}}]{hellieretal91-1}
{Hellier}, C., {Mason}, K.~O., \& {Mittaz}, J.~P.~D. 1991, MNRAS, 248, 5P

\bibitem[{{Hilditch} {et~al.}(1996){Hilditch}, {Harries}, \&
  {Hill}}]{hilditchetal96-1}
{Hilditch}, R.~W., {Harries}, T.~J., \& {Hill}, G. 1996, MNRAS, 279, 1380

\bibitem[{{Homer} {et~al.}(2006){Homer}, {Szkody}, {Henden}, {Chen}, {Schmidt},
  {Fraser}, \& {West}}]{homeretal06-2}
{Homer}, L., {Szkody}, P., {Henden}, A., {Chen}, B., {Schmidt}, G.~D.,
  {Fraser}, O.~J., \& {West}, A.~A. 2006, AJ, 132, 2743

\bibitem[{{Hong} {et~al.}(2009){Hong}, {van den Berg}, {Laycock}, {Grindlay},
  \& {Zhao}}]{hongetal09-1}
{Hong}, J.~S., {van den Berg}, M., {Laycock}, S., {Grindlay}, J.~E., \& {Zhao},
  P. 2009, ApJ, 699, 1053

\bibitem[{{Horne}(1985)}]{horne85-1}
{Horne}, K. 1985, MNRAS, 213, 129

\bibitem[{{Horne} {et~al.}(1993){Horne}, {Welsh}, \& {Wade}}]{horneetal93-1}
{Horne}, K., {Welsh}, W.~F., \& {Wade}, R.~A. 1993, ApJ, 410, 357

\bibitem[{{Kim} {et~al.}(2005){Kim}, {Andronov}, {Park}, \&
  {Jeon}}]{kimetal05-1}
{Kim}, Y.~G., {Andronov}, I.~L., {Park}, S.~S., \& {Jeon}, Y. 2005, A\&A, 441,
  663

\bibitem[{{King} \& {Lasota}(1991)}]{king+lasota91-1}
{King}, A.~R. \& {Lasota}, J.~P. 1991, ApJ, 378, 674

\bibitem[{{King} \& {Wynn}(1999)}]{king+wynn99-1}
{King}, A.~R. \& {Wynn}, G.~A. 1999, MNRAS, 310, 203

\bibitem[{{Knigge}(2006)}]{knigge06-1}
{Knigge}, C. 2006, MNRAS, 373, 484

\bibitem[{{Knigge} {et~al.}(2011){Knigge}, {Baraffe}, \&
  {Patterson}}]{kniggeetal11-1}
{Knigge}, C., {Baraffe}, I., \& {Patterson}, J. 2011, ApJS, 194, 28

\bibitem[{{Kruszewski} \& {Semeniuk}(1993)}]{kruszewski+semeniuk93-1}
{Kruszewski}, A. \& {Semeniuk}, I. 1993, Acta Astron., 43, 127

\bibitem[{{Landi} {et~al.}(2009){Landi}, {Bassani}, {Dean}, {Bird}, {Fiocchi},
  {Bazzano}, {Nousek}, \& {Osborne}}]{landietal09-1}
{Landi}, R., {Bassani}, L., {Dean}, A.~J., {Bird}, A.~J., {Fiocchi}, M.,
  {Bazzano}, A., {Nousek}, J.~A., \& {Osborne}, J.~P. 2009, MNRAS, 392, 630

\bibitem[{{Liedahl} {et~al.}(1995){Liedahl}, {Osterheld}, \&
  {Goldstein}}]{liedahl95etal-1}
{Liedahl}, D.~A., {Osterheld}, A.~L., \& {Goldstein}, W.~H. 1995, ApJ Lett.,
  438, L115

\bibitem[{{Littlefair} {et~al.}(2001){Littlefair}, {Dhillon}, \&
  {Marsh}}]{littlefairetal01-1}
{Littlefair}, S.~P., {Dhillon}, V.~S., \& {Marsh}, T.~R. 2001, MNRAS, 327, 669

\bibitem[{{Littlefair} {et~al.}(2008){Littlefair}, {Dhillon}, {Marsh},
  {G{\"a}nsicke}, {Southworth}, {Baraffe}, {Watson}, \&
  {Copperwheat}}]{littlefairetal08-1}
{Littlefair}, S.~P., {Dhillon}, V.~S., {Marsh}, T.~R., {G{\"a}nsicke}, B.~T.,
  {Southworth}, J., {Baraffe}, I., {Watson}, C.~A., \& {Copperwheat}, C. 2008,
  MNRAS, 388, 1582

\bibitem[{{Mewe} {et~al.}(1986){Mewe}, {Lemen}, \& {van den
  Oord}}]{mewe86etal-1}
{Mewe}, R., {Lemen}, J.~R., \& {van den Oord}, G.~H.~J. 1986, A\&AS, 65, 511

\bibitem[{{Norton} {et~al.}(2008){Norton}, {Butters}, {Parker}, \&
  {Wynn}}]{nortonetal08-1}
{Norton}, A.~J., {Butters}, O.~W., {Parker}, T.~L., \& {Wynn}, G.~A. 2008, ApJ,
  672, 524

\bibitem[{{Norton} {et~al.}(2004){Norton}, {Wynn}, \&
  {Somerscales}}]{nortonetal04-1}
{Norton}, A.~J., {Wynn}, G.~A., \& {Somerscales}, R.~V. 2004, ApJ, 614, 349

\bibitem[{{Patterson}(1994)}]{patterson94-1}
{Patterson}, J. 1994, PASP, 106, 209

\bibitem[{{Patterson} \& {Halpern}(1990)}]{patterson+halpern90-1}
{Patterson}, J. \& {Halpern}, J.~P. 1990, ApJ, 361, 173

\bibitem[{{Patterson} {et~al.}(1998){Patterson}, {Kemp}, {Richman}, {Skillman},
  {Vanmunster}, {Jensen}, {Buckley}, {O'Donoghue}, \&
  {Kramer}}]{pattersonetal98-3}
{Patterson}, J., {Kemp}, J., {Richman}, H.~R., {Skillman}, D.~R., {Vanmunster},
  T., {Jensen}, L., {Buckley}, D. A.~H., {O'Donoghue}, D., \& {Kramer}, R.
  1998, PASP, 110, 415

\bibitem[{{Patterson} \& {Price}(1981)}]{pattersonetal81-2}
{Patterson}, J. \& {Price}, C.~M. 1981, ApJ Lett., 243, L83

\bibitem[{{Patterson} \& {Thomas}(1993)}]{patterson+thomas93-1}
{Patterson}, J. \& {Thomas}, G. 1993, PASP, 105, 59

\bibitem[{{Patterson} {et~al.}(2004){Patterson}, {Thorstensen}, {Vanmunster},
  {Fried}, {Martin}, {Campbell}, {Robertson}, {Kemp}, {Messier}, \&
  {Armstrong}}]{pattersonetal04-1}
{Patterson}, J., {Thorstensen}, J.~R., {Vanmunster}, T., {Fried}, R.~E.,
  {Martin}, B., {Campbell}, T., {Robertson}, J., {Kemp}, J., {Messier}, D., \&
  {Armstrong}, E. 2004, PASP, 116, 516

\bibitem[{{Pretorius}(2009)}]{pretorius09-1}
{Pretorius}, M.~L. 2009, MNRAS, 395, 386

\bibitem[{{Ramsay}(2000)}]{ramsay00-1}
{Ramsay}, G. 2000, MNRAS, 314, 403

\bibitem[{{Ramsay} {et~al.}(2008){Ramsay}, {Wheatley}, {Norton}, {Hakala}, \&
  {Baskill}}]{ramsayetal08-1}
{Ramsay}, G., {Wheatley}, P.~J., {Norton}, A.~J., {Hakala}, P., \& {Baskill},
  D. 2008, MNRAS, 387, 1157

\bibitem[{{Reimer} {et~al.}(2008){Reimer}, {Welsh}, {Mukai}, \&
  {Ringwald}}]{reimeretal08-1}
{Reimer}, T.~W., {Welsh}, W.~F., {Mukai}, K., \& {Ringwald}, F.~A. 2008, ApJ,
  678, 376

\bibitem[{{Ritter} \& {Kolb}(2003)}]{ritter+kolb03-1}
{Ritter}, H. \& {Kolb}, U. 2003, A\&A, 404, 301

\bibitem[{{Rodr{\'{\i}}guez-Gil}
  {et~al.}(2004{\natexlab{a}}){Rodr{\'{\i}}guez-Gil}, {G{\"a}nsicke},
  {Araujo-Betancor}, \& {Casares}}]{rodriguez-giletal04-1}
{Rodr{\'{\i}}guez-Gil}, P., {G{\"a}nsicke}, B.~T., {Araujo-Betancor}, S., \&
  {Casares}, J. 2004{\natexlab{a}}, MNRAS, 349, 367

\bibitem[{{Rodr{\'{\i}}guez-Gil}
  {et~al.}(2004{\natexlab{b}}){Rodr{\'{\i}}guez-Gil}, {G{\"a}nsicke}, {Barwig},
  {Hagen}, \& {Engels}}]{rodriguez-giletal04-2}
{Rodr{\'{\i}}guez-Gil}, P., {G{\"a}nsicke}, B.~T., {Barwig}, H., {Hagen},
  H.-J., \& {Engels}, D. 2004{\natexlab{b}}, A\&A, 424, 647

\bibitem[{{Roming} {et~al.}(2005){Roming}, {Kennedy}, {Mason}, {Nousek}, {Ahr},
  {Bingham}, {Broos}, {Carter}, {Hancock}, {Huckle}, {Hunsberger}, {Kawakami},
  {Killough}, {Koch}, {McLelland}, {Smith}, {Smith}, {Soto}, {Boyd},
  {Breeveld}, {Holland}, {Ivanushkina}, {Pryzby}, {Still}, \&
  {Stock}}]{romingetal05-1}
{Roming}, P.~W.~A., {Kennedy}, T.~E., {Mason}, K.~O., {Nousek}, J.~A., {Ahr},
  L., {Bingham}, R.~E., {Broos}, P.~S., {Carter}, M.~J., {Hancock}, B.~K.,
  {Huckle}, H.~E., {Hunsberger}, S.~D., {Kawakami}, H., {Killough}, R., {Koch},
  T.~S., {McLelland}, M.~K., {Smith}, K., {Smith}, P.~J., {Soto}, J.~C.,
  {Boyd}, P.~T., {Breeveld}, A.~A., {Holland}, S.~T., {Ivanushkina}, M.,
  {Pryzby}, M.~S., {Still}, M.~D., \& {Stock}, J. 2005, Space Science Reviews,
  120, 95

\bibitem[{{Scaringi} {et~al.}(2010){Scaringi}, {Bird}, {Norton}, {Knigge},
  {Hill}, {Clark}, {Dean}, {McBride}, {Barlow}, {Bassani}, {Bazzano},
  {Fiocchi}, \& {Landi}}]{scaringietal10-1}
{Scaringi}, S., {Bird}, A.~J., {Norton}, A.~J., {Knigge}, C., {Hill}, A.~B.,
  {Clark}, D.~J., {Dean}, A.~J., {McBride}, V.~A., {Barlow}, E.~J., {Bassani},
  L., {Bazzano}, A., {Fiocchi}, M., \& {Landi}, R. 2010, MNRAS, 401, 2207

\bibitem[{{Scaringi} {et~al.}(2011){Scaringi}, {Connolly}, {Patterson},
  {Thorstensen}, {Uthas}, {Knigge}, {Vican}, {Monard}, {Rea}, {Krajci},
  {Lowther}, {Myers}, {Bolt}, {Dieball}, \& {Groot}}]{scaringietal11-1}
{Scaringi}, S., {Connolly}, S., {Patterson}, J., {Thorstensen}, J.~R., {Uthas},
  H., {Knigge}, C., {Vican}, L., {Monard}, B., {Rea}, R., {Krajci}, T.,
  {Lowther}, S., {Myers}, G., {Bolt}, G., {Dieball}, A., \& {Groot}, P.~J.
  2011, A\&A, 530, A6+

\bibitem[{{Smith} \& {Dhillon}(1998)}]{smith+dhillon98-1}
{Smith}, D.~A. \& {Dhillon}, V.~S. 1998, MNRAS, 301, 767

\bibitem[{{Southworth} {et~al.}(2007{\natexlab{a}}){Southworth},
  {G{\"a}nsicke}, {Marsh}, {de Martino}, \&
  {Aungwerojwit}}]{southworthetal07-1}
{Southworth}, J., {G{\"a}nsicke}, B.~T., {Marsh}, T.~R., {de Martino}, D., \&
  {Aungwerojwit}, A. 2007{\natexlab{a}}, MNRAS, 378, 635

\bibitem[{{Southworth} {et~al.}(2006){Southworth}, {G{\"a}nsicke}, {Marsh}, {de
  Martino}, {Hakala}, {Littlefair}, {Rodr{\'{\i}}guez-Gil}, \&
  {Szkody}}]{southworthetal06-1}
{Southworth}, J., {G{\"a}nsicke}, B.~T., {Marsh}, T.~R., {de Martino}, D.,
  {Hakala}, P., {Littlefair}, S., {Rodr{\'{\i}}guez-Gil}, P., \& {Szkody}, P.
  2006, MNRAS, 373, 687

\bibitem[{{Southworth} {et~al.}(2007{\natexlab{b}}){Southworth}, {Marsh},
  {G{\"a}nsicke}, {Aungwerojwit}, {Hakala}, {de Martino}, \&
  {Lehto}}]{southworthetal07-2}
{Southworth}, J., {Marsh}, T.~R., {G{\"a}nsicke}, B.~T., {Aungwerojwit}, A.,
  {Hakala}, P., {de Martino}, D., \& {Lehto}, H. 2007{\natexlab{b}}, MNRAS,
  382, 1145

\bibitem[{{Thorstensen}(1986)}]{thorstensen86-1}
{Thorstensen}, J.~R. 1986, AJ, 91, 940

\bibitem[{{Walker}(1954)}]{walker54-1}
{Walker}, M.~F. 1954, PASP, 66, 230

\bibitem[{{Walker}(1956)}]{walker56-1}
---. 1956, ApJ, 123, 68

\bibitem[{{Warner}(1986)}]{warner86-2}
{Warner}, B. 1986, MNRAS, 219, 347

\bibitem[{{Warner}(1995)}]{warner95-1}
---. 1995, Cataclysmic Variable Stars (Cambridge: Cambridge University Press)

\bibitem[{{Warner} \& {Wickramasinghe}(1991)}]{warner+wickramasinghe91-1}
{Warner}, B. \& {Wickramasinghe}, D.~T. 1991, MNRAS, 248, 370

\bibitem[{{Warner} \& {Woudt}(2009)}]{warner+woudt09-1}
{Warner}, B. \& {Woudt}, P.~A. 2009, MNRAS, 397, 979

\bibitem[{{Witham} {et~al.}(2007){Witham}, {Knigge}, {Aungwerojwit}, {Drew},
  {G{\"a}nsicke}, {Greimel}, {Groot}, {Roelofs}, {Steeghs}, \&
  {Woudt}}]{withametal07-1}
{Witham}, A.~R., {Knigge}, C., {Aungwerojwit}, A., {Drew}, J.~E.,
  {G{\"a}nsicke}, B.~T., {Greimel}, R., {Groot}, P.~J., {Roelofs}, G.~H.~A.,
  {Steeghs}, D., \& {Woudt}, P.~A. 2007, MNRAS, 382, 1158

\bibitem[{{Woudt} \& {Warner}(2003)}]{woudt+warner03-1}
{Woudt}, P.~A. \& {Warner}, B. 2003, MNRAS, 340, 1011

\bibitem[{{Wynn} {et~al.}(1997){Wynn}, {King}, \& {Horne}}]{wynnetal97-1}
{Wynn}, G.~A., {King}, A.~R., \& {Horne}, K. 1997, MNRAS, 286, 436

\bibitem[{{Zhang} {et~al.}(1995){Zhang}, {Robinson}, {Stiening}, \&
  {Horne}}]{zhangetal95-1}
{Zhang}, E., {Robinson}, E.~L., {Stiening}, R.~F., \& {Horne}, K. 1995, ApJ,
  454, 447

\bibitem[{{Zorotovic} {et~al.}(2011){Zorotovic}, {Schreiber}, \&
  {G{\"a}nsicke}}]{zorotovicetal11-1}
{Zorotovic}, M., {Schreiber}, M.~R., \& {G{\"a}nsicke}, B.~T. 2011, A\&A, 536,
  A42

\end{thebibliography}

\clearpage


\begin{deluxetable}{lcccc}
\tablecolumns{5}
\tablewidth{0pc}
\tablecaption{Log of the observations\label{t-obslog}, listing the date
  and UT of the observations, the exposure time, and the number of
  frames. All data were obtained in white light.}
\tablehead{
\colhead{Date} &
\colhead{UT}  & 
\colhead{Exp.(s)} &
\colhead{\# Frames}}
\startdata
2006 Dec 22 & 21:25-23:40 & 10-20 & 79\\            
2006 Dec 23 & 20:58-06:37 & 5-10  & 372 \\
2007 Oct 11 & 02:02-03:51 & 35    & 96 \\
2007 Oct 14 & 01:59-06:23 & 35-45 & 225 \\
2007 Oct 15 & 01:51-05:56 & 40    & 225 \\
2007 Oct 16 & 01:53-06:04 & 45    & 210 \\
2009 Nov 23 & 05:48-12:46 & 120   & 179
\enddata
\end{deluxetable}

%
\begin{deluxetable}{ccccc}
\tablecolumns{5}
\tablewidth{0pc}
\tablecaption{The times of eclipse minima of IPHAS\,J0627. \label{t-eclipse_minima}}
\tablehead{
\colhead{Date} & 
\colhead{Eclipse minima (HJD)} & 
\colhead{Cycle} & 
\colhead{$\mathrm{0-C}$\,(s)} & 
\colhead{References}}
\startdata
2004 Nov 30 & 2453340.507625 & 0    &  13 & \citet{withametal07-1}\\
2004 Dec 02 & 2453342.548045 & 6    &   6 & \citet{withametal07-1}\\
2006 Dec 22 & 2454092.429208 & 2211 & -36 & this work \\
2006 Dec 23 & 2454093.449513 & 2214 & -31 & this work \\
2007 Oct 14 & 2454387.621588 & 3079 &  39 & this work \\
2007 Oct 15 & 2454388.641673 & 3082 &  25 & this work \\
2007 Oct 15 & 2454389.661442 & 3085 & -16 & this work \\
2009 Nov 23 & 2455158.929160 & 5347 &  50 & this work 
\enddata
\end{deluxetable}

%
\begin{deluxetable}{lccc}
\tablecolumns{4}
\tablewidth{0pc}
\tablecaption{Additional IPs with the respect to Fig.\,23 of \citet{gaensickeetal05-1} \label{t-IPs}}
\tablehead{
\colhead{IPs} &
\colhead{$\Porb$\,(h)} & 
\colhead{$\Pspin$\,(s)} & 
\colhead{References}}
\startdata
EI\,UMa          &   6.434 & 745.7  & 1,2 \\
RX\,J2133+5107   &   7.193 & 570.82 & 3 \\ 
SDSS\,J2333+1522 &   1.39  & 2499.6 & 4 \\
IGR\,J0022+6141  &   4.033 & 563.53 & 5 \\ 
IGR\,J19267+1325 &   4.58  & 938.6  & 6 \\
IGR\,J15094-6649 &   5.89  & 808.7  & 7, 8 \\ 
XSS\,J00564+4548 &   2.568 & 470.1  & 8, 9, 10 \\
(= 1RXS\,J005528.0+461143) &  & & \\
V597\,Pup        &   2.6687 & 261.9 & 11 \\
IGR\,J16500-3307 &   3.617  & 571.9 & 7, 8 \\
IGR\,J17195-4100 &   4.005  & 1062  & 7, 8 \\
IRXS\,J165443.5-191620 & 3.7 & 546.66 &  12 \\
IGR\,J08390-4833 &   8      & 1480.8  & 8 \\
IGR\,J18308-1232 &   4.2    & 1820    & 8 \\ 
IGR\,J18173-2509 &   6.6    & 831.7   & 8 \\
IPHAS\,J0627     &   8.16   & 2210.27 & this work \\
\enddata
\tablecomments{(1) In addition, we updated the spin periods of
  V2069\,Cyg and 1RXS\,J0636+3535 to be $\Pspin=743.1$\,s and
  $\Pspin=920$\,s \citep{bernardinietal12-1}, respectively. (2) We
  did not include the confirmed IPs with uncertain \Porb\ determined
  e.g. SDSS\,J144659.95+025330.3 \citep{homeretal06-2},
  Swift\,J0732-1331 \citep{buttersetal07-1}, CXOPS\,J180354.3-300005
  \citep{hongetal09-1}, AX\,J1740.2-2903 \citep{gotthelf+halpern10-1},
  and IP candidates such as V426\,Oph and LS\,Peg \citep[][ and
  references therein]{ramsayetal08-1}. }
\tablerefs{(1) \citet{thorstensen86-1}; (2) \citet{reimeretal08-1};
(3) \citet{bonnet-bidaudetal06-1}; (4) \citet{southworthetal07-1}; (5)
\citet{bonnet-bidaudetal07-1}; (6) \citet{evansetal08-1}; (7)
\citet{pretorius09-1}; (8) \citet{bernardinietal12-1}; (9)
\citet{buttersetal08-1}; (10) \citet{bonnet-bidaudetal09-1}; (11)
\citet{warner+woudt09-1}; (12) \citet{scaringietal11-1}}
\end{deluxetable}

\clearpage

\end{document}